\documentclass[useAMS]{aa}
\usepackage{graphicx,natbib}

\newcommand{\mdot}{\dot{m}}

\def\vecu{{\mathbf{u}\ \ignorespaces}}
\def\mdot{{$\dot{m}$\ \ignorespaces}}

\graphicspath{
{/Users/pierre-olivierpetrucci/Boulot/COM/PAPIERS/MICROQ/cygX1/}
}

\begin{document}

\date{Received .../Accepted ...}

\title{The pertinence of Jet Emitting Discs physics to Microquasars: Application to Cygnus X-1}
\titlerunning{The pertinence of JED physics to Microquasars}
\author{P.O.\,Petrucci\inst{1} \and J.\,Ferreira\inst{1} \and G.\,Henri\inst{1} \and J.\,Malzac\inst{2} \and C.\,Foellmi\inst{1}}

\institute{Laboratoire d'Astrophysique, UMR5571 Universit\'e J. Fourier/CNRS, Observatoire de Grenoble BP53,
 F-38041 Grenoble cedex 9, France \and Centre dÕEtude Spatiale des Rayonnements, Universit\'e de Toulouse [UPS], CNRS [UMR 5187], 9 av. du Col. Roche,  31028 
Toulouse, France 
CESR}

\date{}

\abstract
{The interpretation of the X-ray spectra of X-ray binaries during their hard states requires a hot, optically thin medium. There are several accretion disc models in the literature that account for this aspect. However, none is designed to simultaneously explain the presence of powerful jets detected during these states.}   
{A new quasi-keplerian hot accretion disc solution, a Jet Emitting disc (JED hereafter), which is part of a global disc-jet MHD structure producing stationary super-alfv\'enic ejection, is investigated here. Its radiative and energetic properties are then compared to the observational constraints found in Cygnus X-1. }
{We solve the disc energy equation by balancing the local heating term with advection and cooling by synchrotron, bremsstrahlung and Comptonization processes. The heating term, disc density, accretion velocity and magnetic field amplitude were taken from published self-similar models of accretion-ejection structures. Both optically thin and thick regimes are considered in a one temperature gas supported disc.}
{Three branches of solutions are found possible at a given radius but we investigate only the hot, optically thin and geometrically slim solutions. These solutions give simultaneously, and in a consistent way, the radiative and energetics properties of the disc-jet system. They are able to very well reproduce the global accretion-ejection properties of Cygnus X-1, namely its X-ray spectral emission, jet power and jet velocity. About half of the released accretion power is used to produce two mildly relativistic ($v/c\simeq$ 0.5) jets and for a luminosity of the order of 1\% of the Eddington luminosity, the JED tempŽrature and optical depth  are close to that observed in the hard state  Cygnus X-1.} 
{The accretion and ejection properties of JEDs are in agreement with the observations of the prototypical black hole binary Cygnus X-1. The JED solutions are likely to be  relevant to the whole class of microquasars.}


\maketitle

\section{Introduction}

Powerful ejections are commonly observed in galactic black holes. X-ray binaries show permanent jet radio emission during the so-called hard states \citep{stir01,rem06,don07} 
and superluminal sporadic ejection in intermediate flaring states (\citealt{mir94,han01,fom01b,fen04c})\footnote{A tentative summary of our knowledge of accretion-ejection phenomena can be found here http://www.issibern.ch/teams/proaccretion/}. 

Several models have been put forward to explain the formation of these astrophysical jets. They can be roughly classified in two categories both relying on the existence of an organized large scale magnetic field. The first one taps the accretion energy reservoir of the disc, as first proposed by  \citet{lov76} and \citet{bla76}. Since the seminal paper of  \cite{bla82b}, it is  known that jets can indeed be both accelerated and collimated by the magnetic field threading the disc.  In the second category of models, jets are supposed to mainly tap the rotational energy of the black hole thanks to the so-called Blandford-Znajek process (hereafter BZ process, \citealt{bla77}). In this model, accretion serves only to bring in and maintain the magnetic field in the black hole ergosphere.

The fact that the jet power observed in some X-ray binary systems  (the so-called microquasars \citealt{mir98}) seems remarkably  independent of the nature of the compact objects, namely white dwarf, neutron star or black hole (e.g. \citealt{kor06,kor08,tud09}) 
  argues for an ejection process independent of the central object. This seems to be confirmed by the lack of correlation between jet power or jet velocities and measured values of black hole spins \citep{fen10}. Although all these evidences do not constitute a definitive proof, it severely weakens the scenario where powerful jets would be powered by the BZ process. 

There have been numerous 2D and 3D relativistic magnetohydrodynamic (hereafter MHD) numerical simulations of accretion flows around black holes (e.g. \citealt{mck04,kom07,haw06,bec08,mckin09}).
Starting from a geometrically thick donut pervaded by its own weak poloidal or toroidal magnetic field, simulations end up with accretion onto the black hole with outflowing material along opened field lines. Accretion is due to the onset of the magneto-rotational instability or MRI (see \citealt{bal03} and references therein) 
that leads to the formation of an almost geometrically thin disc in the innermost radii. 
A Poynting-dominated funnel jet is formed in the black hole polar regions in good agreement with the BZ process.  
This electromagnetic (no mass) funnel wind is surrounded by a mass-dominated outflow concentrated in a hollow and relatively thin cone at roughly 40 degrees inclination. Ejection is both thermally and magnetically driven and is only present if the initial magnetic field had a vertical component \citep{bec08}. 
Although these simulations look very promising, it is still not clear how they compare to the observed large scale jets. Indeed, it is not clear yet whether such flows will be self-confined further out and if the mass and energy fluxes are in agreement with observational constraints. 
 
On the other hand, following the \citet{bla82b} pioneering work, it has been realized that jets could extract a significant fraction of the underlying disc angular momentum and accretion power. Under such circumstances, jets would not be a mere epiphenomenon of accretion but possibly its main driver. The whole disc structure should then be revisited by taking into account the disc-jets interrelation. 
This was done within a self-similar Ansatz, allowing to solve the full set of dynamical MHD equations without neglecting any term \citep{fer95, fer97, cas00a, cas00b, fer04}. It was found that the structure of  these Jet Emitting Discs (hereafter JED)  is indeed deeply modified, with dynamical properties significantly different from those of a standard accretion disc or an ADAF. As a consequence, the radiative properties of a JED may be also very different. Noticeably, the main results of these steady-state self-similar studies have been confirmed by 2D numerical MHD simulations done with the VAC code \citep{cas02,cas04}, the FLASH code \citep{zan07} and the PLUTO code \citep{tze09}. \\

The purpose of this paper is to detail the dynamical, energetic and radiative properties of Jet Emitting Discs around compact objects and to compare them with those estimated from the observations of the prototype of black hole binaries: Cygnus X-1.

The structure of the paper is as follows. We start by recalling the main characteristics  of a JED in Sect. \ref{jedchar}. We  also present in Sect. \ref{jedsol} the method used to compute the JED thermal equilibrium.  We then compare, in Sect. \ref{cygx1sect} the model expectations with the observational constraints known in Cyg X-1. Concluding remarks are discussed in Sect. \ref{discussion}.

\section{Jet Emitting Discs}
 \label{jedchar}

{Jet Emitting Discs have been originaly studied by  \cite{fer93a} in their Magnetized Accretion Ejection Structures model (MAES herafter). This model has been developed so as to treat 
consistently both the accretion disk and the jet it generates. The 
idea is the same as in earlier studies of magneto-centrifugally 
launched disk winds \citep{bla82b}. However, in 
the MAES, the solution starts from the midplane of the resistive 
MHD disk and evolves outwards in the ideal MHD wind/jet. 
This differs drastically from other studies where the disk was 
only treated as a boundary condition, hence forbidding any 
precise quantiÞcation of the effect of the MHD wind on the disk. 
          
It would be lengthy but also irrelevant to present the MAES 
model in great detail in this paper. Many papers have dealt 
with the subject, from both analytical and numerical point of 
views, and we refer the reader to these papers for further details (e.g. \citealt{fer97,cas00a,cas00b,fer02,fer04,cas02,cas04,zan07,com08}). 
Instead, we give hereafter the few key assumptions and elements of the model that are important to our work.}

\subsection{Main assumptions}
The goal of these studies was to find out the conditions for steady state jet formation and to relate jet properties (mass loss, power, kinematics, collimation) to the underlying disc properties.
The MAES calculations were done under several assumptions that are discussed below: (1) a self-similar Ansatz; (2)  an alpha-prescription for the transport coefficients; (3) the presence of a large scale, organized vertical $B_z$ field.

 (1) Self-similarity
 enables to solve the full set of MHD equations {\em without ignoring any term}. The only simplification that must be made is in the energy equation, since microphysics (within cooling terms) is not self-similar. Thus, given some approximation (i.e. isothermal or adiabatic magnetic surfaces), a computed solution is a real solution of the full set of equations. In our case here, such a solution is obtained by crossing the MHD critical points of the flow. On the other hand, self-similarity introduces unavoidable biases in the jet collimation properties  (see e.g. \citealt{fer97} for more details).

(2) The disc is assumed to be turbulent so that, within our mean field approach, some prescription must be made to mimic turbulence. Following \citet{sha73}, an alpha prescription for the transport coefficients, namely viscosity and magnetic diffusivity, is employed.  It has been shown recently that MRI gives rise to a turbulent magnetic transport that seems to behave like a resistivity, with an effective magnetic Prandtl number of order unity \citep{gua09,lesu09}. 
It turns out that JED solutions are found to be close to the marginal stability limit, but always MRI unstable \citep{fer95, fer97}. Note that there is a whole range of other possible MHD instabilities when the fields are close to equipartition \citep{kep02, blok05, blok07}. 
Whether magnetically driven turbulence can indeed be described with local alpha prescriptions is still an open question \citep{beck09}. Full 3D global simulations are thus required to assess this assumption, but the huge span in space-time scales are such that this is still out of reach of modern computers.   

(3) The last assumption, namely the presence of a large scale magnetic field in the inner regions of accretion flows, is still an open and highly debated question (e.g. \citealt{lub94a,lub94b,cao02,bis07,roth08}).

It has been recently argued however that fields could be advected inward in a standard accretion disc thanks to the quenching of  the MRI at the disc upper layers \citep{roth08,lov09}. This picture has been given some support with recent 2D MHD simulations of accretion-ejection structures \citep{mur10}. GRMHD simulations also suggest that a vertical large scale magnetic field is indeed needed in order to produce powerful jets \citep{beck08}. 
Thus while the controversy is certainly not closed, the assumption of a large scale magnetic field threading the disc appears quite plausible.

\subsection{Physical properties}
\label{sec_physchar}

Throughout the paper, we use $R$ for the cylindrical radius, $M$ for the black hole mass and $\dot M_a(R)$ for the disc accretion rate at the radius $R$. The dimensionless radius will be $r=R/R_g$ where $R_g= GM/c^2$, mass $m = M/M_\odot$ and disc accretion rate $\dot m=\dot M_a c^2/L_{Edd}$ with $L_{Edd}$ the Eddington luminosity. 

\subsubsection{Radial profiles}

Since a JED undergoes mass loss, the disc accretion rate is written as
\begin{equation}
\dot{M}_a(R) = \dot{M}_{a,out} \left (\frac{R}{R_{out}}\right )^\xi
\label{mdot}
\end{equation}
where $\xi$ measures the local disc ejection efficiency \citep{fer93a} and $\dot{M}_{out}$ the accretion rate at the disc outer radius $R_{out}$. A standard accretion disc is recovered when $\xi=0$. The ejection efficiency  $\xi$ is equivalent to the $p$ exponent used in ADIOS models \citep{bla99}. But, in strong contrast to the latter, it is not {\it assumed} but computed as function of the disc parameters as a trans-Alfv\'enic regularity condition (see \citealt{fer97} for more details). For isothermal \citep{fer97} and adiabatic \citep{cas00a} magnetic surfaces, $\xi$ is found to vary in a narrow interval around $0.01$. When the disc upper layers are heated, mass loss can be enhanced with $\xi$ reaching typically 0.1 or slightly more \citep{cas00b}. Discs with larger mass losses do not give rise to stationary super-Alfv\'enic jets. 

{The disc vertical equilibrium depends on the balance between, on one side, gravity and magnetic compression due to both radial and toroidal field components and, on the other side, the vertical pressure (gas + radiation) gradient. We nevertheless use as a proxy for the real scale height the hydrostatic value, namely $P= \rho_o \Omega_K^2 H^2$ evaluated at the disc midplane, where $P= P_{gas} + P_{rad}$: this greatly simplifies our expressions and produces an overestimation of at most a factor 2 \citep{fer95}. With this caveat in mind, the disc aspect ratio {in a gas supported disc} is $\varepsilon= H/R= c_s/v_K$, namely the ratio of the isothermal sound speed to the Keplerian speed $v_K=\Omega_K R$. Given these definitions, the radial profiles of the accretion velocity $u_o$, particle density $n= \rho_o/m_p$, gas pressure $P_{gas}$ and vertical magnetic field $B_z$ in the JED midplane are given by     }
\begin{eqnarray}
u_o &=& - u_r = m_s c_s = m_s \, \varepsilon\, \Omega_K R \label{eq1} \\
n &=& \frac{\dot M_a(R)}{4 \pi m_p \Omega_K R^3 m_s \varepsilon^2} \\
P_{gas} &=& \rho_o c_s^2 = \frac{\dot M_a(R) \Omega_K}{4 \pi R m_s} \\
B_z &=&  \left ( \frac{\mu}{m_s} \right )^{1/2} \sqrt{ \frac{\mu_o \dot M_a(R) \Omega_K}{4 \pi R}}\label{eq2} 
\end{eqnarray} 
where we introduced two dimensionless parameters, the sonic Mach number $m_s= u_o/c_s$ and the disc magnetization\footnote{In a gas supported disc, $\mu= 2/\beta$ where $\beta$ is the well known plasma beta parameter. }
$\mu= \mu_o^{-1}B_z^2/P$. 

\subsubsection{JED parameter space}
\label{jedpar}
The previous expressions (\ref{eq1})-(\ref{eq2}) are common to all power law, nearly Keplerian gas supported disc solutions. What distinguishes JEDs from other models are the values taken by the parameters $\xi, m_s, \mu$ and $\varepsilon$. 

\begin{itemize}
\item The value of $\varepsilon$ is fixed by the resolution of the JED energy equation (see Sect. \ref{jedsol}).

\item $\mu$ and $\xi$ are provided by the resolution of the full set of MHD equations when properly dealing with the regularity conditions at the slow magnetosonic and Alfv\'enic critical points, respectively. We already discussed the values achieved for $\xi$.  
For isothermal or adiabatic magnetic surfaces, steady-state ejection is possible only for a field smaller than but close to equipartition \citep{fer95, fer97, cas00a}. More precisely, the range of allowed magnetizations in isothermal models goes from 0.3 to 0.8, with an average value around 0.6 (see examples of JED parameter values in Fig~\ref{ap1}). This can be understood in the following way. The magnetic field is vertically pinching the accretion disc so that a quasi-static vertical equilibrium is obtained only thanks to the kinetic pressure support. As a consequence, the field cannot be too strong. But on the other hand, the field must be strong enough to accelerate efficiently the plasma at the disc surface so that the slow-magnetosonic point is crossed smoothly.  

\item {  In a JED, the sonic Mach number can be rewritten under the form 
\begin{equation}
m_s=\alpha_v\epsilon+2q\mu
\label{eq:ms}
\end{equation}
where $\alpha_v\epsilon$ denotes the effect of standard transport and $2q\mu$ 
the specific contribution of the magnetic torque. 
The magnetization $\mu=B^2/\mu_0 P$ measures the strength of the magnetic field in the disk and  $q= \mu_o J_r h/B_z$ is the normalized radial electric current density flowing at the disk midplane. This last parameter measures the magnetic shear as it provides an estimate of the toroidal magnetic field component at the disk surface, namely $B_\phi^+ \simeq -q B_z$.

In a SAD, $m_s=\alpha_v\epsilon$, with typically $\alpha_v=10^{-2}$ and $\epsilon=H/R<1$, so that 
the accretion velocity is largely subsonic. However, the situation in a JED is very different since the presence of powerful jets strongly modified the disc angular momentum. A common feature of all solutions found is that the jet torque {\em always} dominates the local turbulent torque, as initially proposed by \citet{pel92b}.
This is because steady-state MAES solutions are generally found close to
equipartition (i.e. $\mu\simeq$ 1 as explained before) and $q$ also of the order of unity
\citep{fer95,fer97}. As a consequence 
JED solutions present high accretion velocities with $m_s\sim 1$. And this is verified by all isothermal and adiabatic solutions (Fig. \ref{ap1}).}


\end{itemize}

In order to simplify the analysis carried out in this paper, we will fix hereafter $m_s= 1$ and $\mu= 0.5$. Putting numbers, one gets:
\begin{eqnarray}  
n  & \simeq & 10^{19}\ \, \varepsilon^{-2}\,  \dot m \, m^{-1}\, r^{-3/2} \ \mbox{cm$^{-3}$}\label{eqa}\\ 
P_{gas} &\simeq& 1.5\ 10^{16} \ \,  \dot m\,  m^{-1}\, r^{-5/2}\ \mbox{erg.cm$^{-3}$}\label{eqab} \\
B_z& \simeq& 4.4\ 10^{8} \ \,  \dot m^{1/2}\,  m^{-1/2}\, r^{-5/4}\ \mbox{G}\label{eqaa} 
\end{eqnarray} 
Note that the \citet{bla82b} scaling for the magnetic field is recovered as we used $\xi\ll 1$.

\subsubsection{Global energy budget}

The global energy budget in an accretion-ejection structure is 
\begin{equation}
P_{acc} = P_{JED} + P_{jets} = P_{rad} + P_{adv} + P_{jets}
\end{equation}
where $P_{jets}$ is the total power feeding the jets and $P_{JED}$ is the power dissipated within the JED. More precisely, $P_{JED}$ is the sum of the radiated $P_{rad}$ and advected $P_{adv}$ power. The disc luminosity is therefore the released accretion power minus what is advected onto the black hole and what feeds the jets.   

Given the expression (\ref{mdot}) of the accretion rate, the total accretion power released in a quasi-Keplerian accretion disc writes\footnote{ This expression neglects any input of energy from the central object at the disc inner edge, through turbulence or MHD Poynting flux (see for instance \citealt{cas00b}).}
\begin{eqnarray}
\label{paccjed}
P_{acc} &=&  \frac{GM\dot{M}_a(R_{in})}{2 R_{in}}-  \frac{GM\dot{M}_a(R_{out})}{2 R_{out}}\nonumber\label{dpacc}\\
&=&\frac{GM \dot M_{a,out}}{2R_{in}} \left [ \left (
\frac{R_{in}}{R_{out}} \right )^\xi - \frac{R_{in}}{R_{out}} \right ]\nonumber\\
 &=& \frac{GM \dot M_{a,out}}{2R_{in}} (1 - g)\label{eq:pacc}
\end{eqnarray}
where
\begin{equation}
(1-g)=\left [ \left (
\frac{R_{in}}{R_{out}} \right )^\xi - \frac{R_{in}}{R_{out}} \right ]
\label{eq:f}
\end{equation}
namely $P_{acc}= \eta\,  \dot m\, L_{Edd}$ with an accretion efficiency $\eta= (1-g)/2r_{in}$. { The usual expression for the non relativistic accretion efficiency is recovered for $\xi=0$. In the following, we will use $r_{in}=6$ for a Schwarzschild black hole.}\\

{The power $P_{adv}$ that is advected onto the central object along with the accreting material scales as $P_{adv} \propto \varepsilon^2 P_{acc}$. It is thus negligible in a thin disc. } $P_{jets}$ is the sum of the total (kinetic, potential and internal) energy flux advected by the outflowing plasma and the MHD Poynting flux leaving the disc surfaces. Since we are mainly interested in powerful jets, the latter contribution dominates i.e. 
\begin{equation}
P_{jets} \simeq P_{MHD} \simeq  - 2\int_{R_{in}}^{R_{out}}  \Omega_* R \frac{B_\phi^+ B_z}{\mu_o}2 \pi R dR 
\label{eqpjet}
\end{equation}
where $B_{\phi}^+$ is the toroidal component of the magnetic field at the disc surface and $\Omega_*$ is the rotation speed of the magnetic surfaces\footnote{A steady-state jet is a bunch of axisymetric magnetic surfaces nested around each other and in differential rotation. Each magnetic surface is put into rotation by the underlying disc, so that $\Omega_* \simeq \Omega^+(r)$, where $\Omega+(r)$ is the angular rotation at the disc surface. Note that this velocity can be significantly smaller than that at the disc midplane.}. Using the disc angular momentum conservation, one gets (see Appendix \ref{apa})
\begin{equation}
P_{jets}\simeq b P_{acc}
 \label{eq:pjet}
 \end{equation}
{where $b$, the jet power parameter, is known for a given MHD solution. Note that in the above derivation, it has been assumed that the fraction of the released energy that is transferred to the jets is a constant throughout the disc. That is not necessarily true. Indeed, self-similar models (where $b$ is by definition a true constant) published in the literature provide $b$ ranging from $\sim$0.5 to almost unity. It turns out that this value depends on the disc aspect ratio but its actual dependence remains to be looked for. As a first step and for the sake of clarity in this paper, we will use it as an independent parameter. }

Consequently, given the above expressions, the total JED power $P_{JED}$ and JED luminosity $P_{rad}$ are given by
\begin{eqnarray}
P_{JED} &=& P_{acc}-P_{jets}=(1-b) P_{acc} \label{eq:jed}\\
P_{rad} &=& P_{JED} - P_{adv} = (1-b) P_{acc} - P_{adv} \label{eq:rad}
\end{eqnarray}
The jet power parameter $b$ is thus a crucial parameter as it controls the power sharing between the jet and the disc. Since it is generally found in the range 0.5-0.99 (see Fig. \ref{ap1}), JEDs considered here are engines converting accretion power into ejection power with a high efficiency. To get the fraction of the power $P_{rad}$ that is actually radiated away, we need however to compute the JED thermal balance. This is explained in the next section.

\section{Disc thermal balance}
\label{jedsol}
The equation governing the internal energy of the accretion flow writes
\begin{equation}
q_{heat}= q_{rad} + \underbrace{P\nabla .\vecu+\nabla U.\vecu}_{q_{adv}} 
\label{eqbalance}
\end{equation}
where $q_{heat}$ is the heating power density, $q_{rad}$ the sum of all radiative cooling terms, $P$ the total plasma pressure, $\vecu$ the flow velocity, $U$ its internal energy and $q_{adv}$ the advection term. These different terms are explicited below. Equation (\ref{eqbalance}) is equivalent to Eq.~(\ref{eq:rad}) but expressed locally in the JED.

Note that in Eq.~(\ref{eqbalance}) we neglect any turbulent energy transport (such as convection for instance) and assume a { Maxwellian} one temperature  plasma. This last point could be a crude assumption at low accretion rates ($\dot m < 10^{-3}-10{-4}$). However, since we are mainly interested in this paper to apply our calculations to Cyg X-1, whose accretion rate is of the order of a few \% of the Eddington rate, this assumption is expected to be quite reasonable (see e.g. \citealt{mal09}).

\subsection{Heating and advection terms}
Equation~(\ref{eq:jed}) provides us directly the radial JED volumetric heating rate in a disc ring of height $2H$, radius $R$ and extent $dR$:
\begin{eqnarray}
 q_{heat}(R)&=&\frac{1}{4\pi RH}\frac{dP_{JED}}{dR}\\
 &=&\frac{(1-b)}{4\pi RH}\frac{dP_{acc}}{dR}=(1-b) \frac{GM \dot M_{a}}{8\pi HR^3}\\
 &\simeq& 1.5\times 10^{21}(1-b)\frac{\dot{m}}{\epsilon m^2r^4}\ \ \mbox{ erg.s$^{-1}$.cm$^{-3}$}
 \label{qheatjed}.
\end{eqnarray}
On the other hand, the advection term is identified as
\begin{equation}
q_{adv}=P\nabla \cdot \vecu + \nabla U.\vecu = \frac{1}{\gamma-1}\left [ \gamma P\nabla \cdot \vecu+ \vecu.\nabla P \right ]
\label{qadvinit}
\end{equation}
where $\gamma=5/3$ is the gas adiabatic index. Assuming that the flow poloidal velocity is dominated by its radial component $u_r$ (provided by Eq.~(\ref{eq1})) and after some algebra, this expression becomes 
\begin{eqnarray}
q_{adv} &= & -\frac{m_p c^3m_s\epsilon^3n }{R^{5/2}}\left ( 4\frac{\partial\ln\epsilon}{\partial\ln R}-1\right ) \nonumber \\
 & \simeq &-3\times 10^{21}\frac{\epsilon\dot{m}}{m^2r^4}\left ( 4\frac{\partial\ln\epsilon}{\partial\ln r}-1\right ) \mbox{ erg.s$^{-1}$.cm$^{-3}$}
\label{qadv}
\end{eqnarray}
The comparison between Eq.~(\ref{qheatjed}) and (\ref{qadv}) shows that advection becomes important only for large aspect ratios, namely $\epsilon>0.1$. Note also that, as already examplified by \cite{yua01,yua03} in ADAF, the advection term may change sign and become negative, playing then the role of a heating term for the flow. Equation (\ref{qadv}) shows that this should occur when the exponent in the radial dependency of $\epsilon$ becomes larger than $1/4$.

\subsection{Radiative cooling term}
We strictly follow \cite{esi96} for the computation of the different radiative cooling rates.  In the optically thin regime, the total cooling rate is simply
\begin{equation}
q_{rad}=q_{C,sync}+q_{C,brem}
\end{equation}
where $q_{C,sync}$ and $q_{C,brem}$ correspond respectively to the synchrotron+ comptonized synchrotron and bremsstrahlung+comptonized bremsstrahlung cooling rates. Since JED solutions may also be optically thick,  we use the generalized radiative cooling formula used by \cite{nar95a}, and adapted from the initial work of \cite{hub90}, that bridges the optically thin and thick cases:
\begin{equation}
q_{rad}=\frac{4\sigma T^4/H}{1.5\tau+\sqrt{3}+(4\sigma T^4/H)(q_{C,sync}+q_{C,brem})^{-1}}
\label{qcool}
\end{equation}
where $\tau=\tau_{es}+\tau_{abs}$ is the disc optical depth in the vertical direction, $\tau_{es}$ being the Thomson optical depth and $\tau_{abs}=(H/4\sigma_BT^4)(q_{C,sync}+q_{C,brem})$ the optical depth for absorption. At high energies, pair creation provides another means of cooling. But an a posteriori check showed that this effect remains always small and has thus been neglected.

\subsection{Existence of three solutions}
\begin{figure}
\includegraphics[height=\columnwidth,angle=-90]{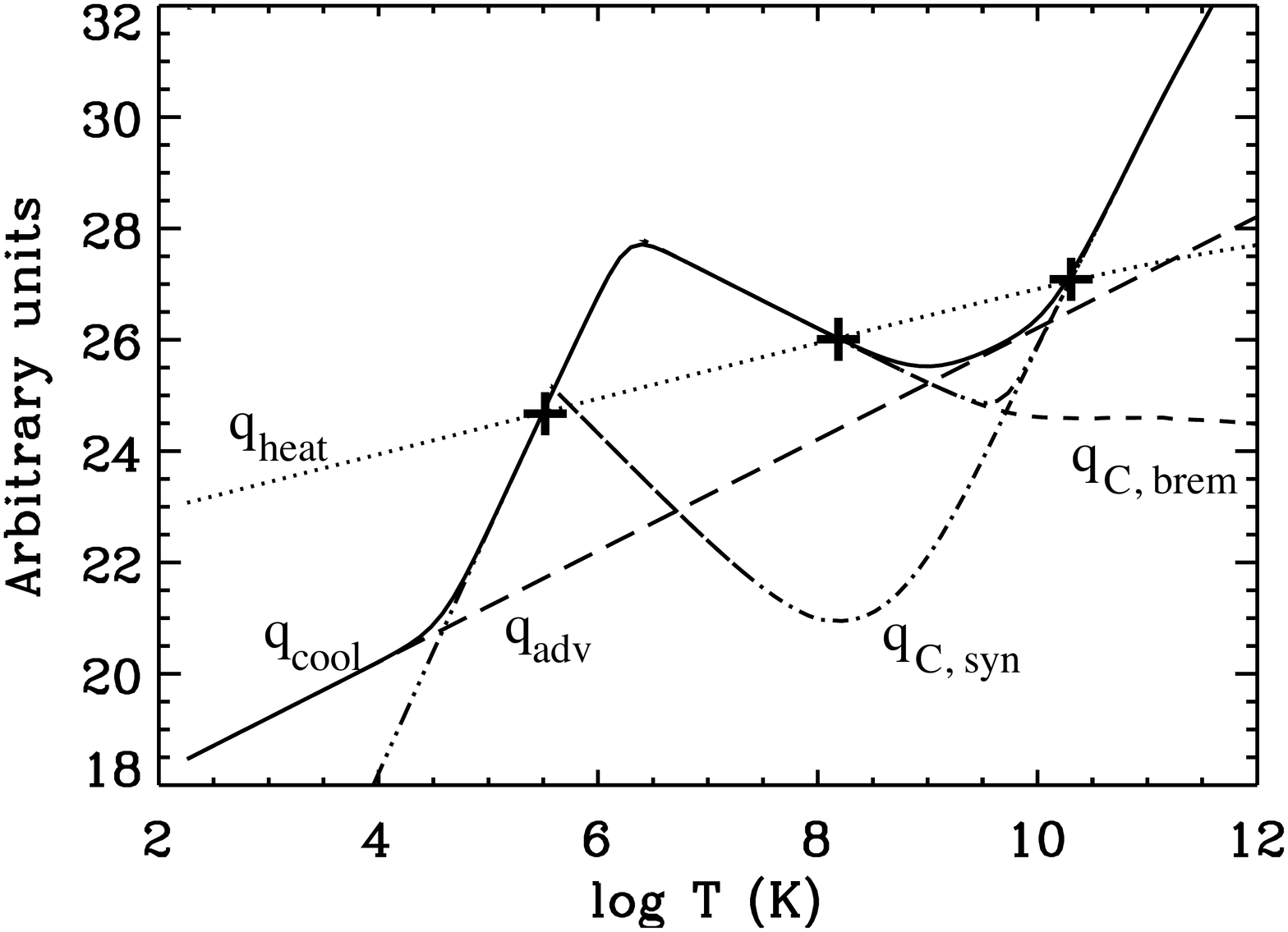}
\caption{Thermal balance of a Jet Emitting Disc at $r=6$ for  \mdot=0.001 and $m=10$. The underlying JED structure has a jet power parameter $b=0.5$ and an ejection efficiency $\xi= 0.1$. 
The dotted line is the heating term $q_{heat}$, the dot-dot-dot-dashed line the radiative cooling term $q_{rad}$  and the long dashed line is the advective term $q_{adv}$. The comptonized Bremsstrahlung $q_{C,brem}$ (short dashed) and the comptonized synchrotron $q_{C,sync}$ (dot-dashed) radiative terms are also shown. The solid line corresponds to $q_{rad}+q_{adv}$ and the solutions of Eq.~(\ref{eqbalance}) are indicated by crosses.}
\label{tfig}
\end{figure}

Figure ~(\ref{tfig}) shows the different heating, cooling and advection terms as function of the temperature for a given black hole mass, JED radius and accretion rate. In a gas supported one-temperature disc, the midplane temperature is given by  
\begin{equation}
T= \frac{m_pc^2}{2k}\epsilon^2\, r^{-1} = 5.4 \times 10^{12}\epsilon^2\, r^{-1}\mbox{ K}
\label{eqtemp}
\end{equation}
The resolution of the thermal equilibrium (Eq. \ref{eqbalance}) gives three branches of solutions, indicated by crosses on the figure 
The hottest one corresponds to the JED solution discussed in this paper { (a more detailed discussion of the other solutions is postponed to a future work)}. It is a hot, optically thin, thermally and viscously stable disc solution that behaves radiatively in a similar way than the LHAF solutions studied by \cite{yua01} (see also \citealt{yua06} for a discussion on hot one-temprature accretion flows). The major difference concerns the underlying dynamics of our JED solutions, that self-consistently include the presence of powerful self-collimated jets.

\begin{figure}
\includegraphics[height=\columnwidth,angle=-90]{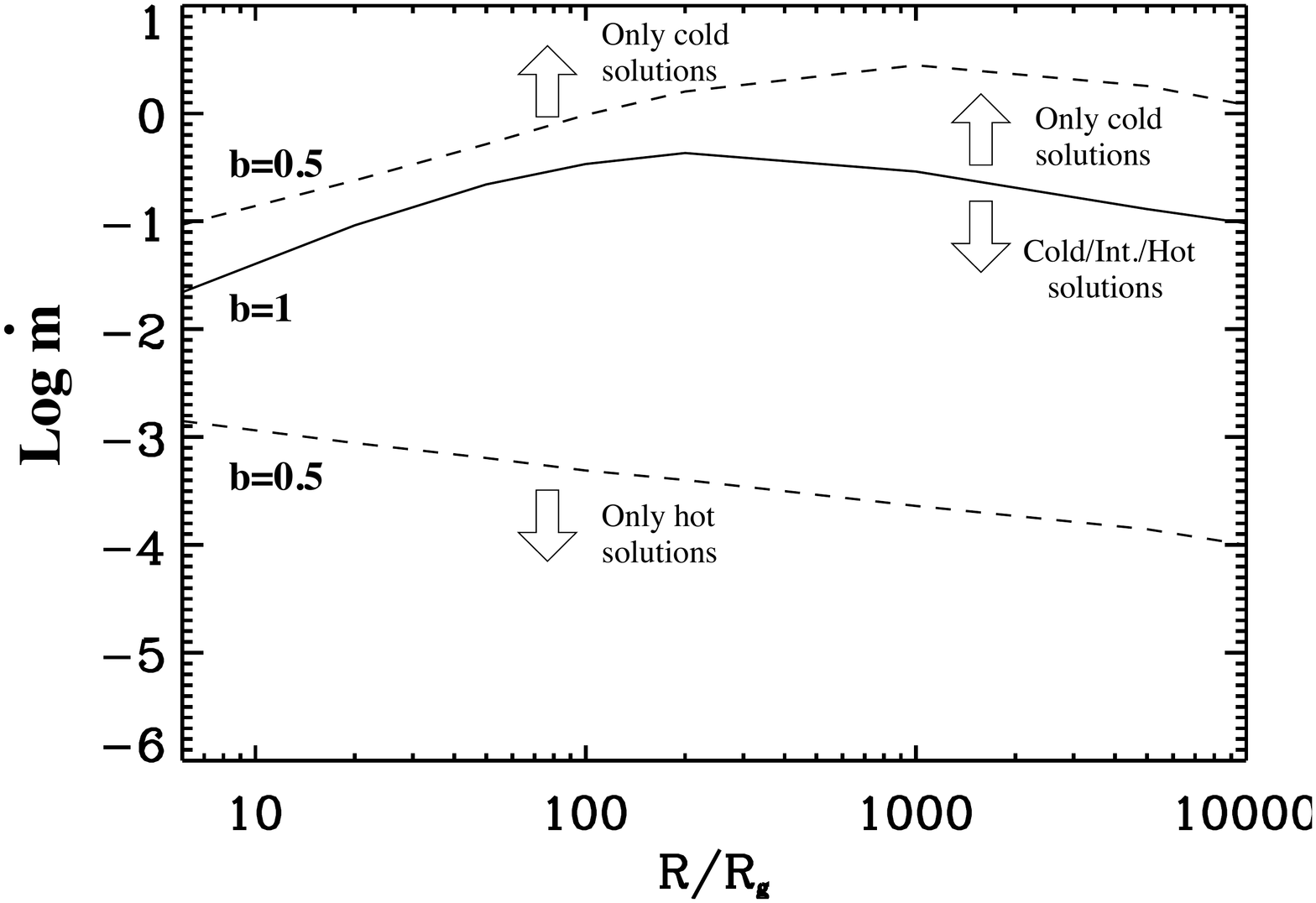}
\caption{JED solution domain in the \mdot-$r$ plane.  The solid line, corresponding to a jet power parameter $b=1$, separates this plane in 2 regions:  the top one where only the cold solution is valid and the bottom one where the 3 solutions exist. The region where only a hot solution is valid exists at very low accretion rate not shown on this figure. The limits for $b=0.5$ are plotted in dashed lines. In this case the plane is divided in 3 parts with a region, at low accretion rate, where only the hot solution exists.}
\label{res6}
\end{figure}

Fig. \ref{res6} also indicates the different domains of JED solutions in the \mdot-$r$ plane and for the two extreme values of the jet power parameter $b$, 0.5 and 1. In the latter case, a cold branch exists in the whole JED whatever the value of the accretion rate while the hot branch  is only valid below a critical accretion rate that depends on the JED radius (solid line in Fig. \ref{res6}). For $b=0.5$, the parameter space for the cold branch is now limited to accretion rates $> 10^{-3}-10^{-4}$ (bottom dashed line) while the upper limit for the hot branch has shifted to slightly larger accretion rate (top dashed line) compared to $b=1$, reaching \mdot of a few for JED radius larger than $\sim$100 $R_g$. A more detailed discussion of the JED solutions is presented in F10.

\begin{figure*}[t!]
\begin{tabular}{cc}
\includegraphics[height=\columnwidth,angle=-90]{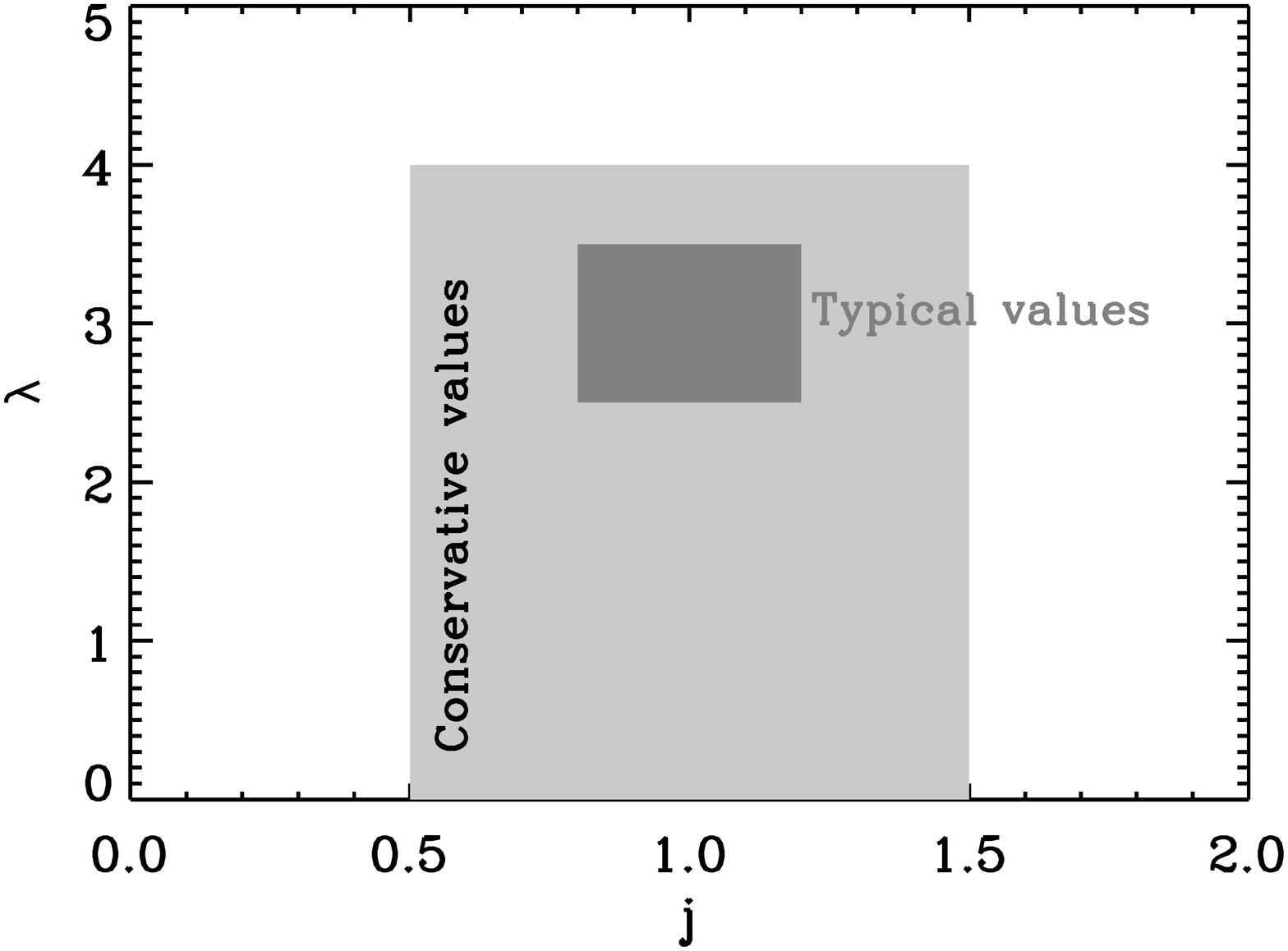}
&\includegraphics[height=\columnwidth,angle=-90]{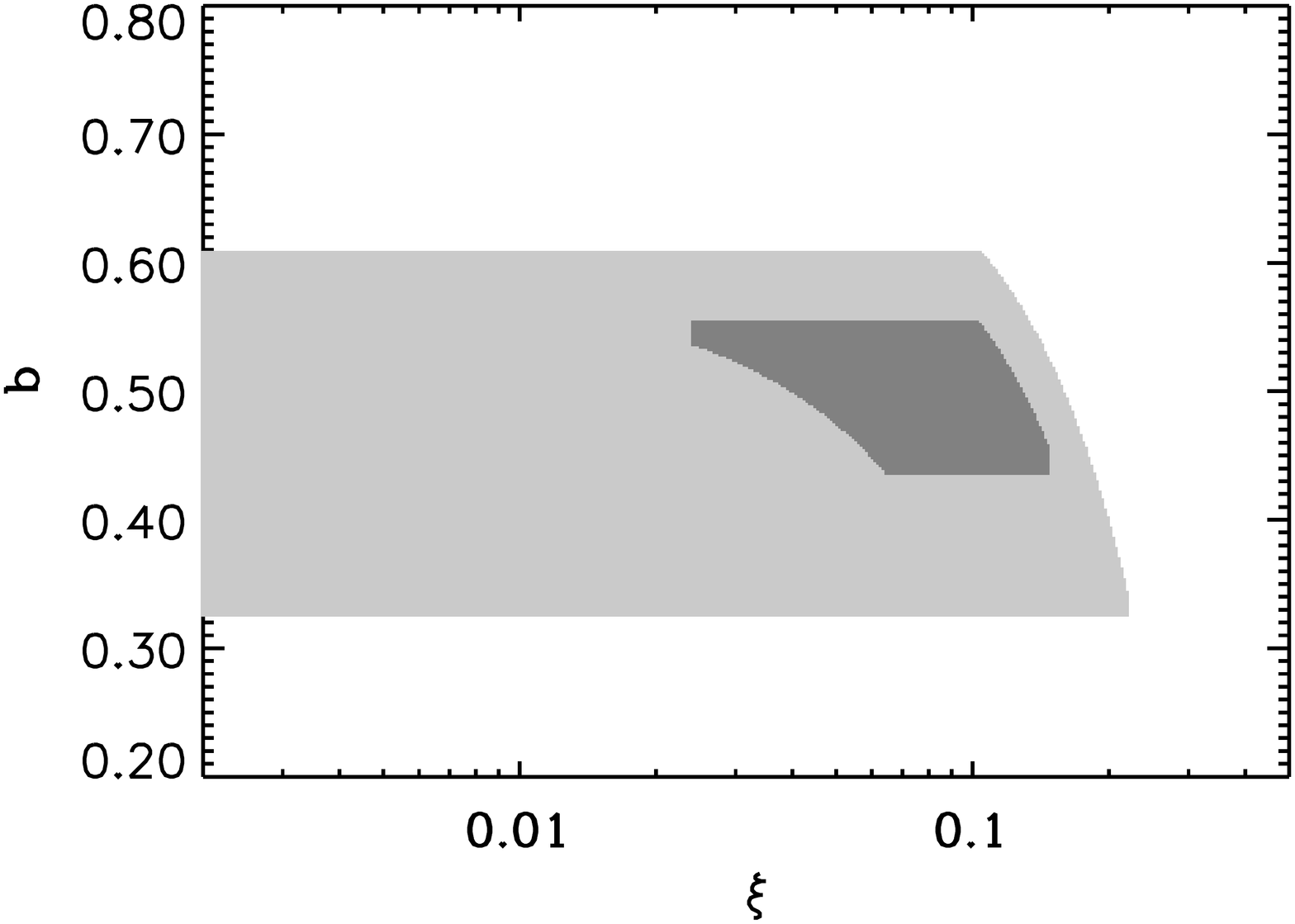}\\
\end{tabular}
\caption{{\bf Left:} Observational parameter space of $\displaystyle j=\frac{P_{jets}}{L_{h}}$ and $\displaystyle \lambda=\frac{L_s}{\dot{M}_s}\frac{\dot{M}_h}{L_h}$. The grey area corresponds to conservatives values of these two parameters while the dark grey one corresponds to more commonly adopted ones (see M09 for more details). {\bf Right:} 
Corresponding parameter space in the $b-\xi$ JED parameter plane.
}
\label{bxiplane}
\end{figure*}


\section{Comparison to Cygnus X-1}
\label{cygx1sect}
\subsection{Observational constraints}

The energetics of the jets and the X-ray corona of Cygnus X-1 have been investigated recently by Malzac et al. (\citeyear{mal09b} herafter M09). It is one of the best observed black-hole X-ray binary composed of a massive star and a black hole of about 10 solar masses. It is also one of the rare BHB for which a jet was directly observed in the hard state \citep{stir01}. But more importantly deep radio observations of the field of CygX-1 resulted in the discovery of a 
shell-like structure which is aligned with the resolved radio jet \citep{gal05b}. Assuming that this large-scale structure was inflated by the inner radio jet, its kinetic power was estimated to be of the order of the bolometric X-ray luminosity of the binary \citep{gal05b,rus07}. With these constraints it is possible to deduce a  rough estimate of the ratio $j=P_{jets}/L_{h}$ of the total jet kinetic power to the typical X-ray luminosity in the hard state (M09)
\begin{equation}
 0.45\leq j=\frac{P_{jets}}{L_h} \leq 1.5
\label{eqj}
\end{equation}
Like most X-ray binaries, the bolometric luminosity of Cyg X-1 does not change dramatically during the transition to the soft state (see e.g. \citealt{fro01,zdz02,mal06,wilm06}).
The observed luminosity jumps reach at most  a factor $L_s /L_h\simeq 3-4$  \citep{zdz02}. The transition to soft state being likely triggered by an increase in mass accretion rate, the ratio of the soft to hard radiative efficiencies can be safely upper limited by
\begin{equation}
\lambda=\frac{L_s}{\dot{M}_s}\frac{\dot{M}_h}{L_h} \leq 4
\label{eqlambda}
\end{equation}

It is noteworthy that there are also several estimates of the terminal jet velocity of Cyg X-1. Indeed, based on the absence of detection of the counter jet, \cite{stir01} give a lower limit on the bulk velocity of the radio jet of $V_{\infty}/c > 0.6$. Similar considerations and the lack of response of the radio emission on short time-scales led \cite{glei04} to constrain the jet velocity in the range $0.4 < V_{\infty}/c < 0.7$. \cite{Ibra07}  find a similar result, $0.3 < V_{\infty}/c < 0.5$, by modeling the super-orbital modulation observed in X-ray and radio bands. This gives another independent constraint that will be very useful while confronting models to observations.

\subsection{Model comparison}
These observational constraints on $j$ and $\lambda$ can be easily translated into constraints on our JED parameters $b$ and $\xi$. From Eqs.~(\ref{eq:pjet}) and (\ref{eq:rad}) and assuming that advection is negligible (which is verified for JEDs at $\dot{m}$ of a few \% of the Eddington accretion rate, as expected for Cyg X-1) we obtain
\begin{equation}
j=\frac{b}{1-b}
\label{eq:j}
\end{equation}
{The soft state is free of ejection (i.e. $\xi=0$) so that $L_s = \eta_s \dot m_s\, L_{Edd}$, with $\eta_s \simeq 1/2r_{in}$. The hard state  is characterized by $L_h = (1-b) \eta_h \dot m_h\, L_{Edd}$, with $\eta_h \equiv (1-g)/2r_{in}$ where $g$ in the hard state is defined in Eq.~(\ref{eq:f}) and then depends on $\xi$ and the JED radial extent $R_{out}/R_{in}$. Within this framework, we obtain 
\begin{equation}
\lambda=\frac{1}{(1-b)(1-g)}.
\label{eq:lambda}
\end{equation}

The left panel of Fig. (\ref{bxiplane}) displays the domain in the observed parameter space $j-\lambda$ allowed by the observations. Conservative values are shown in grey whereas most plausible ones, i.e. $j \simeq 1$ and $\lambda\simeq 3$ \citep{mal09b}, are in dark grey. The right panel shows the same constraints but translated into our JED parameter space $\xi-b$, assuming $R_{out}/R_{in}=100$. 
{This value is comparable to what could be deduced from observations (e.g. \citealt{fab89,you01,tor05}) if the standard accretion disc inner edge is directly assimilated to the JED outer radius as proposed in \cite{fer06a}. Note that this figure depends only slightly on this parameter with a smooth increase of the $\xi$ upper limit with decreasing  $R_{out}/R_{in}$. 
%
We can thus safely conclude that  $\xi$ has to be {\it lower} than $\sim$0.1 and $b$ ranging between 0.4 and 0.6 to be consistent with Cyg X-1 data. \\  


\begin{figure}
\includegraphics[height=\columnwidth,angle=-90]{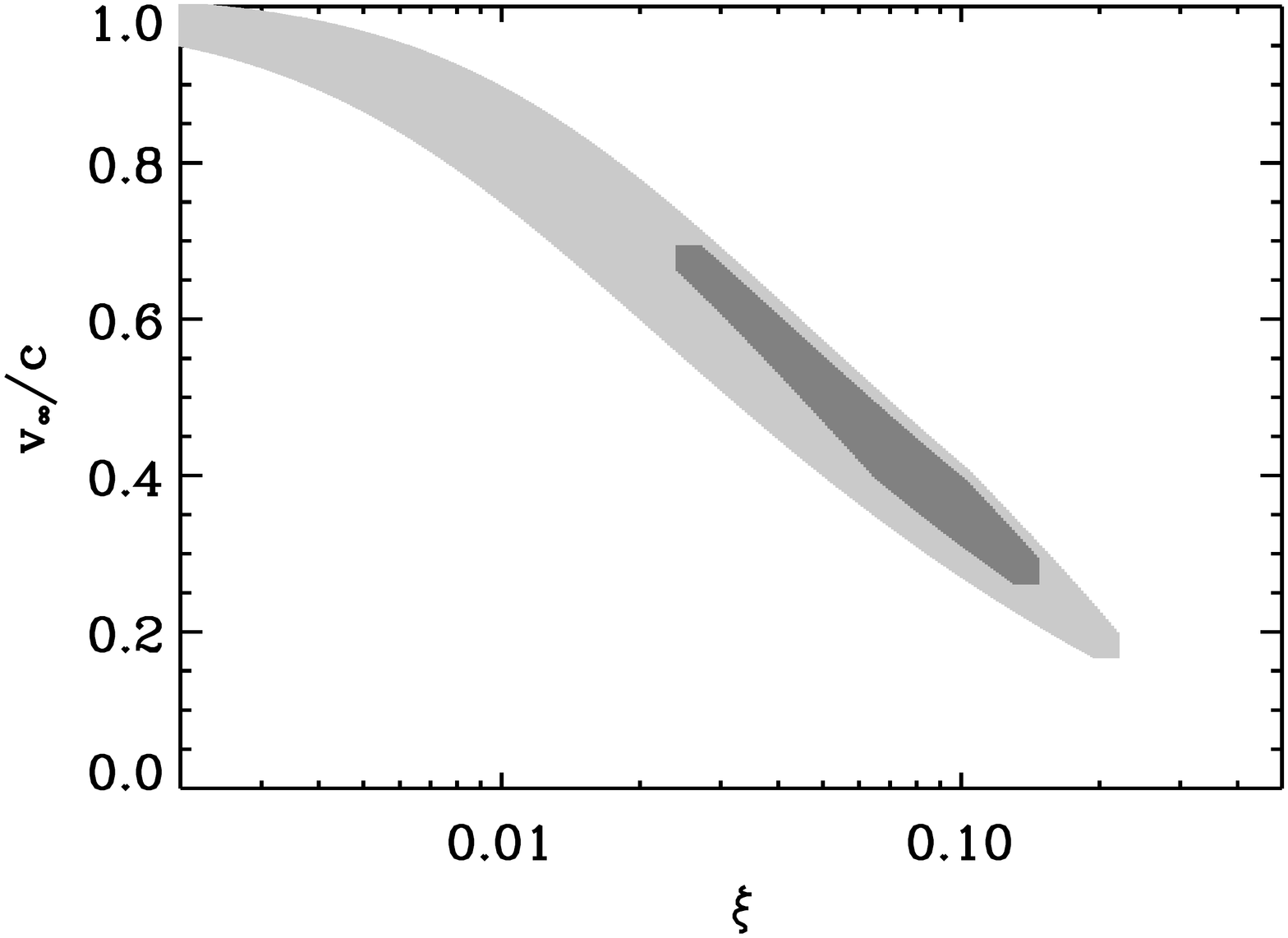}
\caption{Average terminal jet velocity versus the disc ejection efficiency $\xi$ in a JED, with $\dot M_a \propto R^\xi$. The grey and dark grey areas correspond to the regions that agree with the observational constraints reported in Fig.~(\ref{bxiplane}).}
\label{betainffig}
\end{figure}

Now, the jet terminal velocity can be also estimated. Since the jets of Cyg X-1 are not highly relativistic, we can assume that the jet power $P_{jets}$ is asymptotically dominated by the kinetic energy of the ejected matter. This provides an average terminal bulk Lorentz factor
\begin{equation}
\langle\gamma_{\infty}\rangle \simeq1+\frac{P_{jets}}{\dot{M}_{jets}c^2}
\label{jetvel}
\end{equation}
where the total mass loss rate $\dot{M}_{jets}$ is such that 
\begin{equation}
f= \frac{\dot{M}_{jets}}{\dot M_a(R_{out})} = 1-\left ( \frac{R_{in}}{R_{out}}\right )^{\xi} \simeq \xi \ln  \frac{R_{out}}{R_{in}}
\end{equation}
the last equality holding only for small $\xi$. Combining this expression with Eq.~(\ref{eq:pjet}), we finally obtain
\begin{equation}
\langle\gamma_{\infty}\rangle  \simeq 1+  \eta \frac{b}{f} = 1+ \frac{b}{2r_{in}}\frac{(1-g)}{f}
\end{equation}
The range of average jet velocities $V_{\infty}/c$ consistent with the values of $b$ and $\xi$ needed to agree with Cyg X-1 observations can then be computed from this last expression. Figure (\ref{betainffig}) shows $V_{\infty}/c$ as function of the disc ejection efficiency $\xi$. The grey area corresponds to the conservative values used in Fig.~(\ref{bxiplane}), the dark grey area to adopted ones. These last ones provide a range in $V_{\infty}/c$ between 0.3 and 0.6 in good agreement with observations. This is remarkable as it arises from another, independent observational constraint.\\

\begin{figure}
\includegraphics[height=\columnwidth,angle=-90]{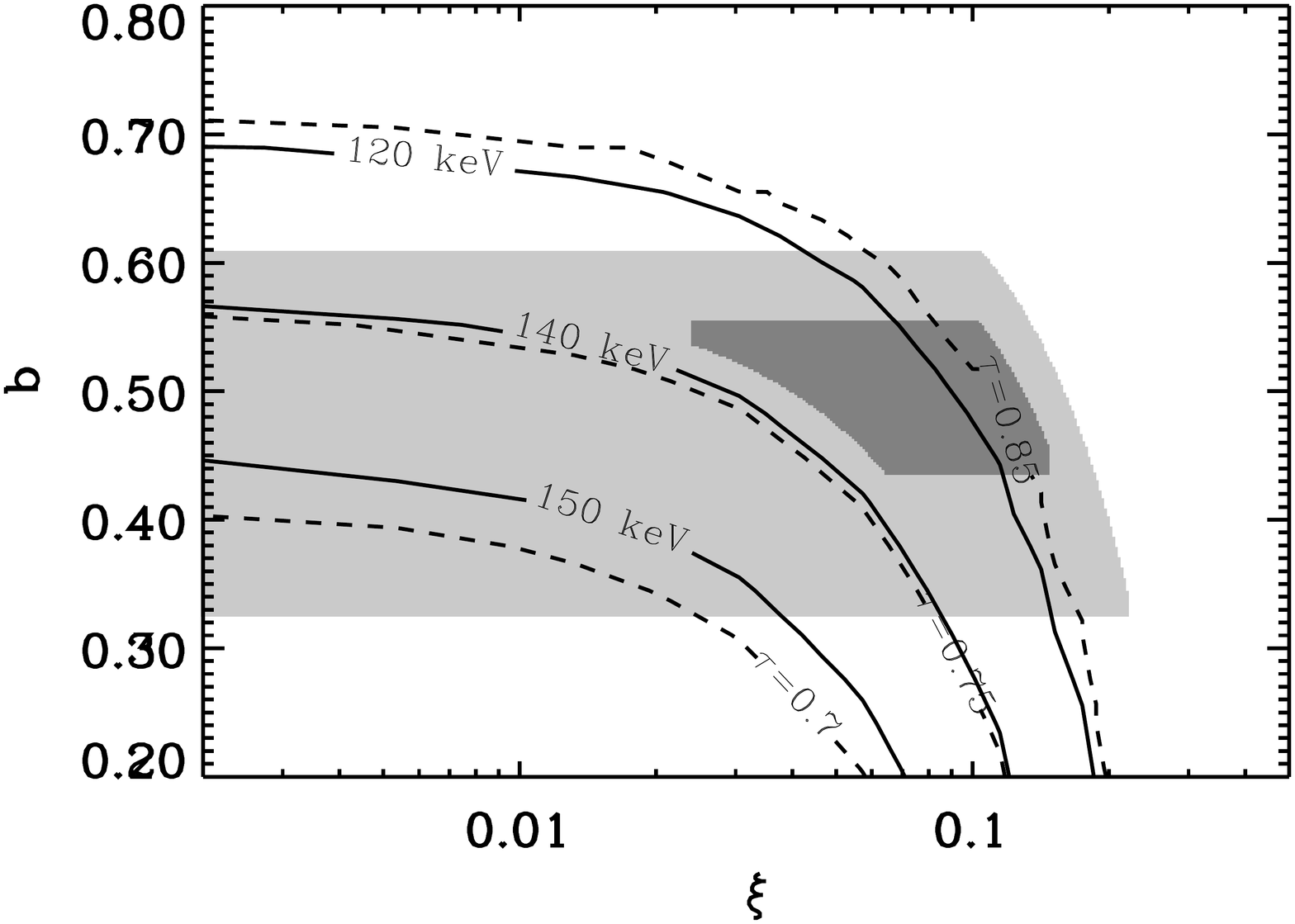}
\caption{Contour plots of the JED temperature (solid lines) and optical depth (dashed lines) in the $\xi-b$ plane at a distance of 10 $R_G$ from a 10 $M_\odot$ black hole. The ratio of the outer to inner JED radii is taken equal to 100 and the total accretion rate for the computation of the disc thermal balance is fixed to 1\% of $L_{Edd}$, i.e. the typical luminosity of Cyg X-1. The grey and dark grey areas correspond to the regions that agree with the observational constraints reported in Fig.~(\ref{bxiplane}). 
}
\label{temptaufig}
\end{figure}

Our treatment of the disc thermal balance allows us to give an estimate of the expected temperature and optical depth within a JED. Contours of these two quantities at a radius of 10 $R_G$ are shown in Fig.~(\ref{temptaufig}), the solid lines representing the temperature while the dashed lines the optical depth. The black hole mass is taken equal to 10 $M_{\odot}$ and the total luminosity is about 1\% of the Eddington luminosity. This corresponds to the typical luminosity of Cyg X-1 close to the hard-to-soft state transition. The ratio of the outer to inner JED radii is fixed to 100 but Fig.~(\ref{temptaufig}) does not strongly depend on this parameter. Interestingly, the temperature is of the order of 120-140 keV and the optical depth $\tau\simeq 0.7-0.9$, which are not so far, given the simplicity of our thermal balance computation, from the values deduced from sophisticated fits of Cyg X-1 in the hard state \citep{salvo01,gier97,fro01,ibra05,cad06,wilm06}.

\section{Concluding remarks }
\label{discussion}

{
Jet Emitting Discs (JED) can be much less radiative than Standard Accretion Discs fed at the same accretion rate. Indeed, JEDs redistribute the released accretion power into three outputs: radiation, advection onto the central object and jet power. Using published self-similar calculations of JEDs, which provided us with the relevant disc parameters, we were able to solve the disc thermal balance around compact objects in a consistent way.  

As in other studies (e.g. \citealt{chen95}), we found that there are usually three branches of solutions at a given radius: (1) a cold solution, optically thick and geometrically thin; (2) an intermediate solution and (3) a hot solution, optically thin and geometrically thicker. Only the cold and hot solutions are both thermally and viscously stable. In this paper,  we focussed on the properties of the hot solution as it corresponds to the physical situation envisioned during the hard states of X-ray binaries, which are known for harboring jets. We confronted our JED model to observations by comparing the disc X-ray emission, jet power and jet velocity with those derived in Cygnus X-1.\\


We found that these hot JED solutions reproduce well both dynamical and radiative properties of Cygnus X-1. These solutions are characterized by a jet power parameter $b \sim 0.5$ and an ejection efficiency $\xi < 0.1$. 
The first condition arises from the observational fact that, in Cyg X-1, the gravitational power is apparently equally shared between the accretion flow and the jet. This requires discs with aspect ratio quite large, namely $\varepsilon \simeq 0.1$. Such solutions, like our JED ones, are nowadays available 
On the other hand, if $\xi$ were too large the disc accretion rate (scaling as $r^\xi$) at the inner radius during the hard state would be much smaller than that at the outer radius, while there are basically the same during the soft state. This would therefore lead to a ratio of radiative efficiencies much larger than what is currently observed. This upper limit on the local disc ejection efficiency translates into an upper limit of the jet mass loss, namely  $\dot{M}_{jets}/\dot{M}_a(r_{out})\simeq \xi \ln(r_{out} /r_{in} ) <50$\% (for a JED extension arbitrarily chosen to be $r_{out} /r_{in} =100$). At first sight, this may appear not too constraining. However, accretion-ejection dynamics link the jet mass loss to the jet asymptotic velocity. This is the reason why any independent measure of the jet terminal velocity is so important. By comparing the average velocity provided by the models to the values derived from observations, we conclude that any JED model with $\xi$ of a few percent matches all observational constraints for Cygnus X-1 jets (velocity, power, mass loss).\\
}

Concerning the disc itself, we found JED temperatures of 120-140 keV and optical depths $\tau \simeq 0.7-0.9$ for Cyg X-1 parameters. These temperature/optical depth values are  slightly larger/smaller than those generally deduced from observations, namely $T_e\simeq 70-100$ keV and $\tau\simeq 1-3$ \citep{gier97,fro01,ibra05,cad06,wilm06}.
However, our value of the Compton parameter $y\simeq 4\tau\frac{kT_e}{m_ec^2}$ agrees well with the observations. We thus expect a X-ray spectrum with the correct spectral index but with a cut-off (mainly controlled by the temperature) too high. {This could be directly linked to our overly simple treatment of the disc radiative transfer which neglect for example its vertical stratification. }
%
%
One might also argue that the electron particle distribution is not a perfect Maxwellian. Indeed, recent computations of plasma radiative equilibria, including radiative and kinetic processes, show that a non-thermal high energy tail in the particle distribution is generally expected \citep{bel08,mal09}. The high energy photons produced by these particles could create pairs, then increasing the particule density and decreasing the temperature. These different effects will be more precisely studied in a fortcoming paper.\\

The presence of a non-thermal high energy tail above the thermal Comptonisation cut-off in the hard state of Cygnus X-1  \citep{mccon02} and GX 339-4 \citep{war02,joi07} suggests that the magnetic field is low in the Comptonizing region. Indeed, this excess is believed to be the signature of a population of non-thermal electrons in the corona.  These high energy electrons produce a self-absorbed synchrotron emission in the IR-optical bands. These soft photons are then Comptonised in the hot plasma. If the synchrotron luminosity is too strong then it is impossible to sustain the relatively high ($\sim$ 100 keV) temperature of the Maxwellian electrons.  In the case of Cygnus X-1, \cite{war01} estimated that the non-thermal tail implies a strongly sub-equipartition magnetic field. Then based on a detailled spectral modelling  \cite{mal09} and \cite{pou09} infer that the ratio of magnetic to radiation energy density $U_{B}/U_{R}<0.3$.  In the framework of our model this would imply a magnetization that is much too small to allow for the production of a jet.

However this constraint on the magnetic field holds only if  the non-thermal tail is  produced inside  the thermal Comptonisation region. A possibility would be that the bulk of the Comptonised emission in the hard state is produced within the JED, while the non-thermal comptonisation component arises from a magnetic  accretion disc corona above and below the standard outer disc.  If so, the magnetic field could be strong in both JED and non-thermal corona.  Then, during spectral state transitions, the non-thermal component would gradually get stronger and stronger as  the transition radius between outer disc and JED moves closer to the black hole. In the soft state the JED disappears and the hard X-ray emission is dominated by a non-thermal Comptonisation in the corona.



\section*{Acknowledgments}

This work has been partly supported by the French National Agency (ANR) through the project "Astro2flots" ANR-05-JCJC-0020 and the GdR PCHE in France. POP also thanks the International Space Science Institute (ISSI) for hospitality.

\appendix
\section{Jet power and \mbox{b} parameter} 
\label{apa}

The power leaving the disc and carried away by the jets (from both face) is $P_{jets} =2 \int \rho \vec u_p E \cdot \vec dS$, where the integration is done along one disc surface from $R_{in}$ to $R_{out}$ and
\begin{equation}
E = \frac{u^2}{2}\,  +\, H\, +\, \Phi_G\, - \Omega_* \frac{R B_\phi}{\eta} 
\label{eq:Bern}
\end{equation}
is the well known Bernoulli integral. It is the sum of all specific energies carried by the jet, namely kinetic, enthalpy, potential and magnetic. In this last term, $\Omega_*$ is the rotation speed of the magnetic surface and $\eta$ is the mass flux per magnetic flux unit. 

In the ideal MHD regime realized in jets, $E$ is an invariant that can be evaluated at the disc surface. In a thin disc ($\varepsilon \ll 1$), it writes $E \simeq \frac{v_K^2}{2} (2\lambda_{BP} - 3)$, where  $\lambda_{BP}$ is the  \citet{bla82b} magnetic lever arm parameter. For the models considered here, $\lambda_{BP}$ varies between 10 and 100 so that the magnetic contribution dominates in Eq.~(\ref{eq:Bern}). The jet power leaving the disc is therefore mostly in the form of an MHD Poynting flux, 
\begin{equation}
P_{jets} \simeq P_{MHD} \simeq  - 2\int_{R_{in}}^{R_{out}}  \Omega_* R \frac{B_\phi^+ B_z}{\mu_o}2 \pi R dR 
\end{equation}
which is exactly Eq.~(\ref{eqpjet}). The toroidal magnetic field at the disc surface scales as $B_\phi^+ = - F_1 \mu_o J_o H$ where $J_o$ is the radial electric current density at the disc midplane, with $F_1 \simeq 0.4-0.6$ for most published solutions. In a JED, where the dominant torque is due to the jets, the disc angular momentum conservation is
\begin{equation}
J_o B_z =  \frac{\rho_o u_o}{R} \frac{\partial }{\partial R} \Omega_o R^2= \frac{1}{2} \rho_o u_o \Omega_o
\end{equation}
where the subscript "o" stands for values evaluated at the disc midplane. Writing $\displaystyle\dot M_a(R) = - 4 \pi R \int_0^H \rho u_r dz = 4 \pi R \rho_o u_o H F_2$, with $F_2$ a number of order unity found to vary between 0.4-0.6 also for most of our solutions, we obtain
\begin{equation}
P_{jets} \simeq 2\int_{R_{in}}^{R_{out}}  dR\, \frac{G M \dot M_a(R)}{2R^2} f_{jet}
\end{equation}
In this expression, $f_{jet}(R)= \displaystyle\frac{F_1}{F_2} \frac{\Omega_*}{\Omega_K} \frac{\Omega_o}{\Omega_K}$ is a priori a function of the radius. However it cannot vary much if ejection is to occur on a large radial extent. Under this assumption, our self-similar scalings apply and we obtain Eq.~(\ref{eq:pjet}) with
\begin{equation}
b \simeq \frac{F_1/F_2}{1-\xi} \frac{\Omega_*}{\Omega_K} \frac{\Omega_o}{\Omega_K} \simeq \frac{\Omega_*}{\Omega_K} \frac{\Omega_o}{\Omega_K} 
\end{equation}
While this quantity is nearly independent of $\xi$, it is clearly a function of the disc thickness, hence its temperature. The rotation of the disc material at the disc midplane scales as $\Omega_o = \Omega_K \sqrt{1 - \frac{5}{2}\varepsilon^2 - p \mu \varepsilon}$, where the second term is due to the radial pressure gradient and the third to the magnetic tension ($ p \sim B_r^+/B_z \sim 1$). Thus, the thicker the disc, the larger the deviation from Keplerian and the smaller $b$. 

But the dominant effect is due to the rotation of the magnetic surfaces $\Omega_*$. Indeed, $\Omega_*$ is defined only in the upper layers of the disc once ideal MHD becomes valid. The disc density has therefore decreased significantly, enhancing thereby the outward effect of the magnetic radial tension. This tends to decrease the angular velocity $\Omega^+$ of the matter and thus of the magnetic surfaces $\Omega_*$ (see for instance the vertical profiles shown in \citealt{fer95}). This behaviour can be also understood with the magnetic torque. This torque is first negative before becoming positive at the disc upper layers. The disc is therefore first spun down and $\Omega$ decreases. 

Numerical solutions obtained with isothermal magnetic surfaces provide $b$ ranging from 0.99 for thin discs with $\varepsilon$ smaller than $\sim 10^{-3}$ to roughly 0.5 for slim discs with $\varepsilon \sim 0.1$.

\begin{figure*}
\begin{tabular}{cc}
\includegraphics[height=\columnwidth,angle=-90]{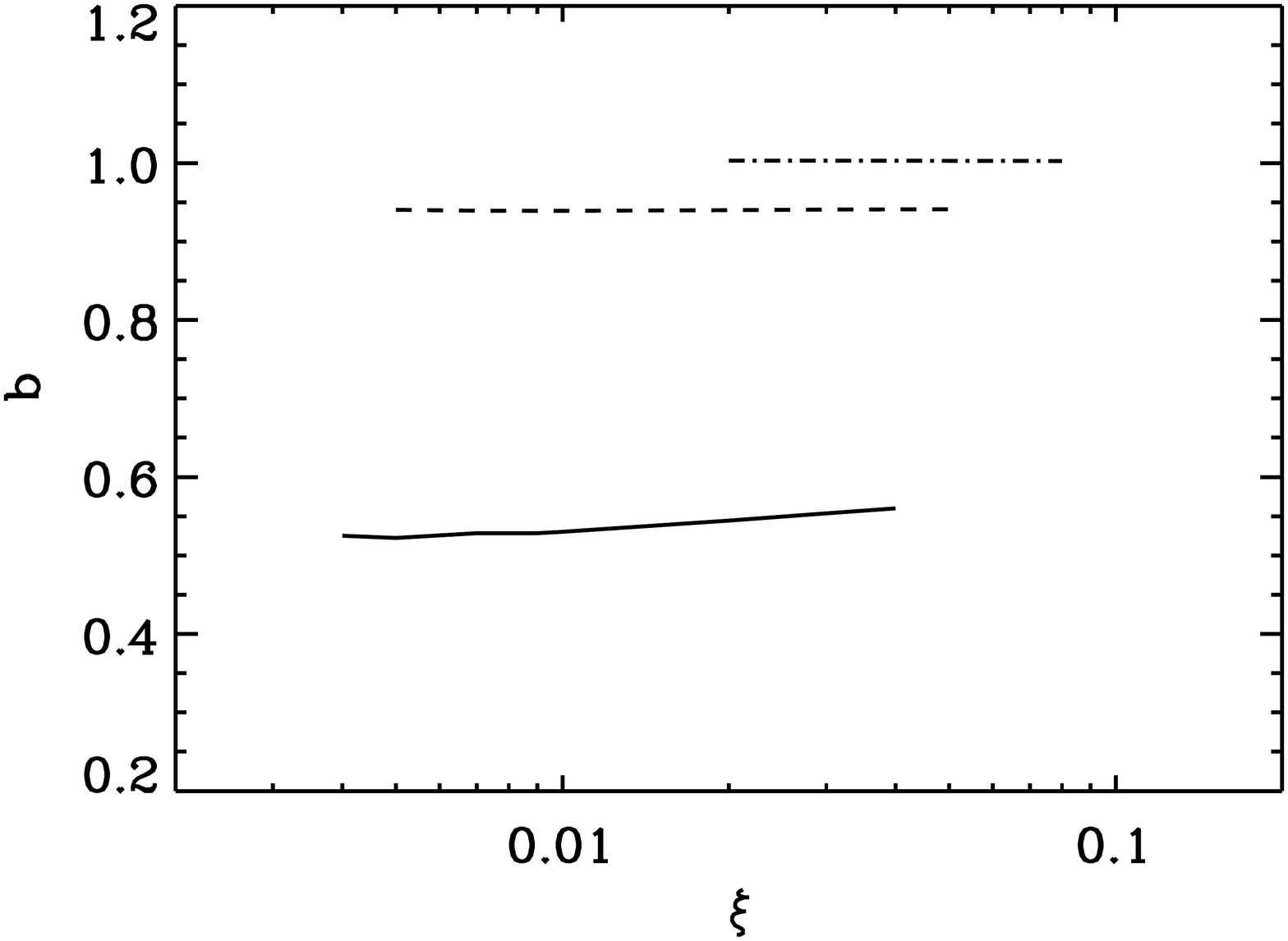} &\includegraphics[height=\columnwidth,angle=-90]{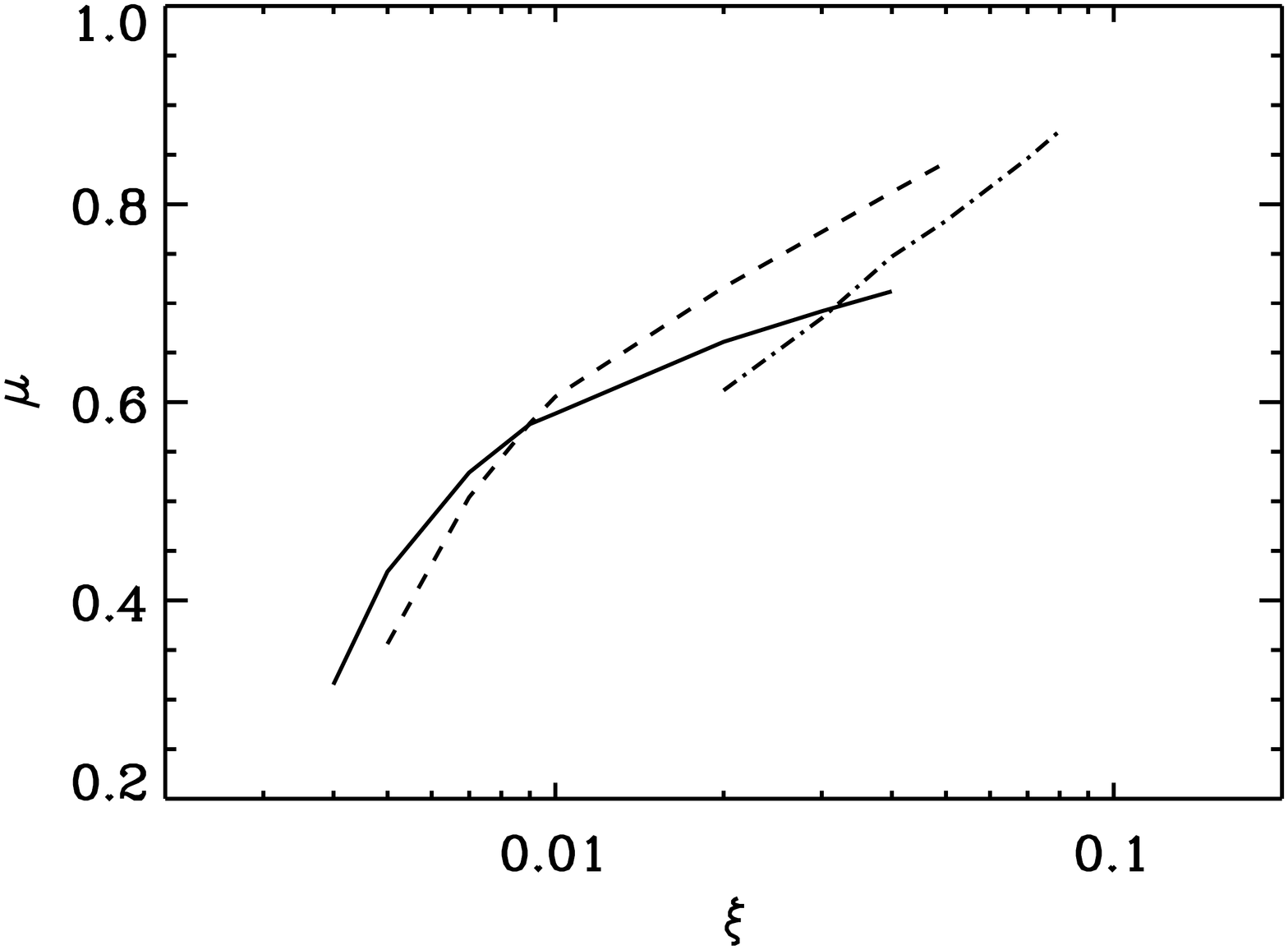}\\
\includegraphics[height=\columnwidth,angle=-90]{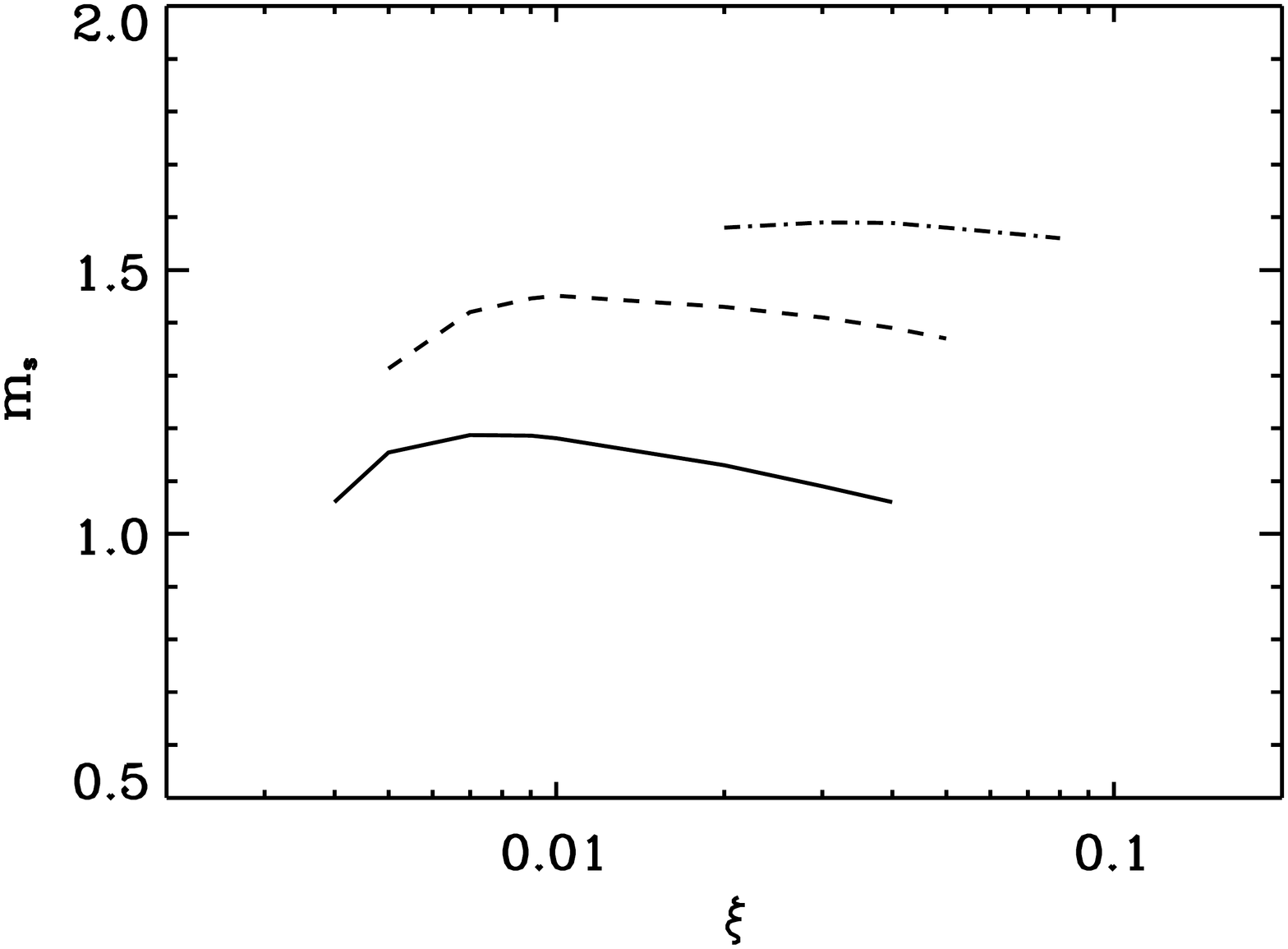} &
\begin{minipage}{\columnwidth}
\vspace*{0cm}
\caption{Parameter space for the ejection efficiency $\xi$, the magnetization $\mu$, the jet power parameter $b$ and the sonic Mach number $m_s$ for the accretion-ejection solutions of \cite{fer97} used in this paper. The lines correspond to different values of the disc aspect ratio $H/R$: solid lines: $H/R=0.1$, dashed lines: $H/R=0.01$, dot-dashed lines: $H/R=0.001$. For this set of solutions, the turbulence level parameter $\alpha_m$ is taken equal to unity. \label{ap1}}
\end{minipage}\\
\end{tabular}

\end{figure*}


\end{document}